\documentclass{article}

\usepackage{tabularx}
\usepackage{array}
\newcolumntype{Y}{>{\raggedright\arraybackslash}X}

\usepackage{microtype}
\usepackage{graphicx}
\usepackage{booktabs}
\usepackage{wrapfig}
\usepackage{subcaption}

\usepackage[ruled,vlined,noend]{algorithm2e}

\usepackage{amsmath,amssymb,amsfonts}
\usepackage{textcomp}
\usepackage{xcolor}
\usepackage{soul}
\usepackage[normalem]{ulem}
\usepackage{float}

\newtheorem{observe}{Observation}

\usepackage[most]{tcolorbox}
\definecolor{light-gray}{gray}{0.95}
\tcolorboxenvironment{thm}{
  center,
  boxsep=1.0pt,
  top=0.0pt,
  bottom=0.1pt,
  width=\linewidth,
  colframe=light-gray,
  colback=light-gray
}

\newcommand{\squishlist}{
\begin{list}{$\bullet$}
  { \setlength{\itemsep}{0pt}
    \setlength{\parsep}{0pt}
    \setlength{\topsep}{0pt}
    \setlength{\partopsep}{0pt}
    \setlength{\leftmargin}{0.7em}
    \setlength{\labelwidth}{0.7em}
    \setlength{\labelsep}{0.2em} } }
\newcommand{\squishend}{
  \end{list}  }

\newcommand{\name}{{CIERA}\xspace}

\usepackage[utf8]{inputenc} 
\usepackage[T1]{fontenc}    
\usepackage{hyperref}       
\usepackage{xspace}         
\usepackage{cleveref}       
\usepackage{url}            
\usepackage{booktabs}       
\usepackage{amsfonts}       
\usepackage{nicefrac}       
\usepackage{microtype}      
\usepackage{xcolor}         

\newcommand{\DEL}[1]{\iffalse #1 \fi}

\title{CIERA: Cross-Iteration Exponent Reuse for Lossless
Allgather in Sharded MoE Training}

\author{%
  Ali Zafar Sadiq \\
  University of Virginia \\
  \texttt{mzw2cu@virginia.edu}
  \and
  Haiying Shen \\
  University of Virginia \\
  \texttt{hs6ms@virginia.edu}
  \and
  Masahiro Tanaka \\
  Anyscale \\
  \texttt{mtanaka@anyscale.com}
}

\begin{document}

\maketitle

\begin{abstract}
In training Mixture-of-Experts (MoE) models, sharded data parallelism partitions each expert’s parameters across GPUs, requiring an Allgather operation to reconstruct the full weight matrix before each layer executes. This communication often dominates iteration time. Prior work often reduces this overhead using lossy compression methods that sacrifice numerical fidelity, while existing lossless methods do not exploit cross-iteration exponent stability. In this paper, we propose Cross-Iteration Exponent Reuse Allgather (CIERA), a lossless, system-aware communication method for sharded MoE training. We observe that after a brief warmup phase, the exponent values of most weights remain unchanged across iterations. Based on this, we cache exponents locally and transmit only the sign and mantissa when exponents are unchanged. The receiver reconstructs the original weights exactly by combining the cached exponents with the received data. Since parameter matrices vary in shape across layers, compression is applied only when it yields net time savings. Moreover, the compression operation is overlapped with both Allgather communication and computation. Our real experiments and large-scale trace-driven simulator show that on OLMoE-1B-7B at 16 GPUs, \name achieves a 3.70× speedup over the lossless baseline and 3.68× over the lossy baseline, projected to reach 4.28× and 4.42× respectively at 128 GPUs, while preserving bitwise-exact parameter reconstruction in all evaluated runs. 

\vspace{-0.15in}
\end{abstract}




\section{Introduction}

Large Language Models (LLMs)~\cite{devlin2019bert,jiang2024mixtral} have become increasingly popular for natural language processing applications such as universal chatbots. As model sizes scale, training these models imposes substantial computational demands. Mixture-of-Experts (MoE) architectures alleviate this overhead by introducing sparsity, activating only a small subset of specialized sub-models (experts) per input token while maintaining model quality~\cite{jiang2024mixtral}. Recent fine-grained MoE designs, such as the DeepSeek series, push this idea further by employing many smaller experts, enabling near-trillion-parameter training without a proportional increase in computational overhead~\cite{liu2024deepseek, dai2024deepseekmoe}.

However, while sparsity reduces computation and improves scalability, it introduces additional communication overhead during training. To distribute a large MoE model across multiple GPUs, sharded data parallelism~\cite{megatroncore_custom_fsdp_2025,megatroncore_moe_api_2025} partitions each expert’s parameters across GPUs. This design keeps tokens local but requires Allgather communication to reassemble weights before computation in the forward pass and aggregates gradients in the backward pass. However, Allgather communication becomes a critical bottleneck in MoE because of its low computation-to-communication ratio, which leaves insufficient computation to overlap and hide communication latency. 


To reduce communication overhead in sharded data parallelism, methods such as ZeRO++~\cite{wang2023zero++} compress weights using block-based quantization, where a block is a small subset of a parameter tensor~\cite{dettmers20218}. Each block is quantized independently with its own scale and rounding factors, making the quantization inherently lossy. 
gZCCL~\cite{huang2024gzccl} compresses data by approximating values within a chosen error bound and encoding the approximated values more compactly before transfer, followed by decompression upon receipt. However, these lossy compression methods introduce numerical error between the original values and the re-constructed ones, degrading training accuracy. On the other hand, lossless methods~\cite{zhou2021designing} insert compression and decompression into communication operations, which can slow training~\cite{zhou2023accelerating}. 



To address these problems, we propose Cross-Iteration Exponent Reuse Allgather (CIERA), a lossless Allgather method for sharded MoE training, motivated by our observations below:
\vspace{-0.05in}

\squishlist

\item[O1:] Allgather accounts for $51.8$--$69.0\%$ of communication time.

\item[O2:] At least 99\% of weights retain unchanged exponents across iterations after a brief warmup.


\item[O3:] Block change rates vary substantially across shard types (attention and routing-gate shards change nearly every iteration, layer norm is consistently stable, and expert Feed-Forward Network (FFN) is stable for some models such as OLMoE but less so for others), and synchronous exponent checking on the critical path costs $25$--$30\%$ of iteration time.




\squishend




\name consists of three components.
First, \textbf{exponent reuse based compression} splits each weight into
exponent, sign, and mantissa, caches the exponent, and sends only sign and
mantissa when the exponent is unchanged. Second, \textbf{benefit-driven selective compression} compresses weights only for shard types where the expected communication savings exceed the compression overhead. Third,
\textbf{computation-communication pipelining} schedules exponent checking
and compression early so they run during preceding compute or communication,
and overlaps receiver-side decompression with Allgather.

We implement \name on top of ZeRO-3~\cite{rajbhandari2020zero} and evaluate it on up to 16 A100 GPUs across six MoE models in BF16 and FP16, achieving $3.70\times$ speedup over ZeRO-3 (lossless) and $3.68\times$ over ZeRO++ (lossy) on OLMoE-1B-7B at 16 GPUs while preserving bitwise-exact parameter reconstruction in all evaluated runs.

\noindent\textbf{Contributions.} 
First, we empirically characterize the Allgather bottleneck and exponent stability in sharded MoE training across multiple MoE models.
Second, we present \name, a lossless, system-aware communication method that selectively compresses only profitable shards and overlaps most compress/decompress work with compute and communication via FX-graph scheduling, leaving only a short decompress on the critical path.
Third, we evaluate \name against ZeRO-3, FSDP, and ZeRO++ on six MoE models in BF16 and FP16, and verify bitwise-exact parameter reconstruction on OLMoE-1B-7B.

\section{Motivation and Observations}
\label{sec:motivation}
We run all experiments on a single server with 4 NVIDIA A100-80GB SXM
GPUs interconnected via NVLink 3.0 with NVSwitch, providing 600~GB/s
bidirectional bandwidth per GPU. The server uses CUDA 12.4, PyTorch 2.7.0, and DeepSpeed 0.17.2.
Unless otherwise stated, all runs use BF16
precision and the Adam optimizer~\cite{kingma2014adam} with the AG News~\cite{zhang2015character}
training dataset and a learning rate (LR) of $10^{-4}$. We use the following four MoE models: OLMoE-1B-7B~\cite{muennighoff2024olmoe} (64 experts per layer),
DeepSeek-MoE-16B~\cite{dai2024deepseekmoe} (64 fine-grained experts),
MiniCPM-MoE-8$\times$2B~\cite{hu2024minicpm} (8 experts per layer), and
Qwen2-57B-A14B~\cite{yang2024qwen2} (28 fine-grained experts). We run 30 warmup iterations, then 1000 subsequent iterations, and report the average iteration time over those 1000 steady-state iterations.


\subsection{Allgather Bottleneck}
\label{sec:mot-allgather}

\begin{wrapfigure}{r}{0.40\columnwidth}
  \vspace{-0.53in}
  \centering
  \includegraphics[width=\linewidth,height=1.2in]{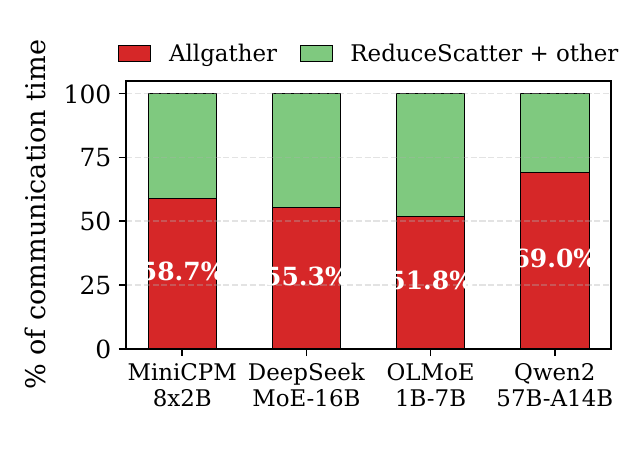} \vspace{-0.3in}
  \caption{Allgather share of communication time.}
  \label{fig:allgather_pct}
  \vspace{-0.2in}
\end{wrapfigure}
The Allgather operation is a significant communication bottleneck in sharded MoE training. Under ZeRO-3 sharding, each weight matrix $W$ is partitioned across GPUs and reassembled on demand by an Allgather before each forward and backward layer~\cite{nccl_collective_operations,deepspeed_zero3_doc}.

Sharded data parallelism has two communication phases every iteration: (1) Allgather, which reconstructs the full parameters from shards during the forward and backward passes, and (2) ReduceScatter, which aggregates gradients after the backward pass. Figure~\ref{fig:allgather_pct} quantifies these time overheads across four MoE models with BF16, $L{=}4$, and $S{=}1024$, where $L$ denotes the number of layers and $S$ denotes sequence length. The figure breaks down profiled GPU time into Allgather, ReduceScatter, and the other communication operations, including AllReduce for gradient updates and layer-wise synchronization. Allgather accounts for $51.8$--$69.0\%$ of total communication time.

\vspace{-0.1in}
\begin{observe}\label{obs:allgather}
Allgather becomes a dominant bottleneck in sharded MoE training especially for large models (Figure~\ref{fig:allgather_pct}).
\end{observe}

\subsection{Exponent Stability}
  \label{sec:exponent-stability}            
            \begin{wrapfigure}{r}{0.62\columnwidth}
  \vspace{-0.80in}
  \centering
  \begin{subfigure}[t]{0.48\linewidth}
    \includegraphics[width=\linewidth,height=1.2in]{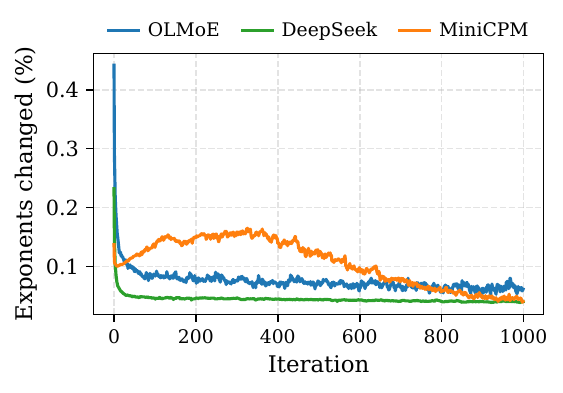}
    \caption{LR $= 10^{-5}$}
    \label{fig:exponent-stability-a}
  \end{subfigure}
  \hfill
  \begin{subfigure}[t]{0.48\linewidth}
    \includegraphics[width=\linewidth,height=1.2in]{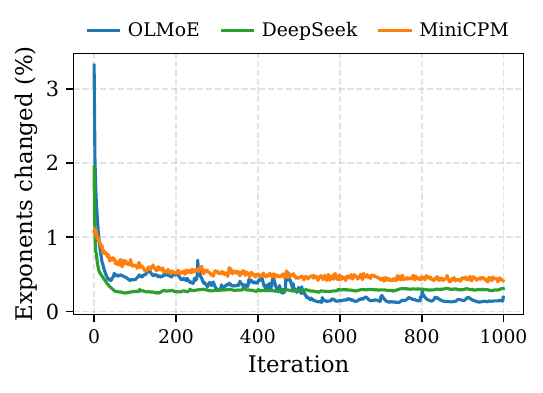}
    \caption{LR $= 10^{-4}$}
    \label{fig:exponent-stability-b}
  \end{subfigure}
  \caption{Exponent change rate during training.}
  \label{fig:exponent-stability}
  \vspace{-0.15in}
\end{wrapfigure}  
 
To check exponent stability, we train OLMoE-1B-7B, DeepSeek-MoE-16B, and
MiniCPM-MoE-8$\times$2B with learning rates $\{10^{-5}$,
$ 10^{-4}\}$ and track
exponent changes at every iteration, as shown in
Figure~\ref{fig:exponent-stability}. At iteration $t$, we extract the
floating-point exponent field of each individual weight value in the weight matrices and
compare it with that in iteration $t{-}1$. Fig.~\ref{fig:exponent-stability-a} shows the exponent change rate over 1000 iterations, which is the
fraction of weight values whose exponent changes between two consecutive iterations. 
Across all three models, after the first 30 iterations of warmup, at
least 99\% of exponent values remain unchanged from one iteration to the
next. Across three models, the exponent change rate stays very low. With $\mathrm{LR}{=}10^{-5}$, it averages $0.04\%\text{ to }0.11\%$ over 1000 iterations and falls to $0.04\%\text{ to }0.06\%$ by iteration 1000. Moreover, at $\mathrm{LR}{=}10^{-4}$, it averages $0.30\%\text{ to }0.51\%$ and falls to $0.19\%\text{ to }0.41\%$ by iteration 1000.

\DEL{For OLMoE-1B-7B at $\mathrm{LR}{=}10^{-4}$, the average change rate is
0.30\% across 1000 iterations, dropping to 0.19\% by iteration 1000. At
$\mathrm{LR}{=}10^{-5}$, the average is 0.07\% across 1000 iterations. DeepSeek-MoE-16B and
MiniCPM-MoE-8$\times$2B show similar stability patterns: across the three models, the change rates are within 0.2\% at $\mathrm{LR}{=}10^{-5}$ and within 1\% at $\mathrm{LR}{=}10^{-4}$.}

To check if the same results hold for other training datasets, we used the OpenWebText pretraining corpus~\cite{gokaslan2019openwebtext}. Here, we observe post-warmup mean exponent change rates of $0.93\%$ (OLMoE), $0.44\%$ (DeepSeek), and $0.62\%$ (MiniCPM) at $\mathrm{LR}{=}10^{-4}$.

\vspace{-0.1in}
\begin{observe}\label{obs:stability}
After the first 30 iterations of warmup, at least $99\%$ of exponent values remain unchanged between consecutive iterations  
(Figure~\ref{fig:exponent-stability}).
\end{observe}

\subsection{Shard-Level vs.\ Block-Level Exponent Reuse}

  \label{sec:shard-vs-block}

\begin{wrapfigure}{r}{0.62\columnwidth}
  \vspace{-0.10in}
  \centering
  \begin{subfigure}[t]{0.48\linewidth}
    \includegraphics[width=\linewidth,height=1.2in]{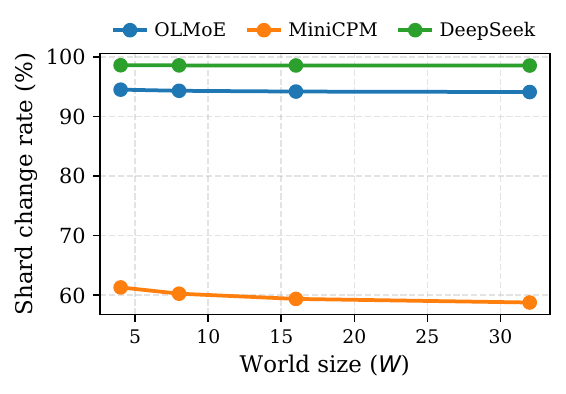}
    \caption{Shard change rate.}
    \label{fig:shard-level}
  \end{subfigure}
  \hfill
  \begin{subfigure}[t]{0.48\linewidth}
    \includegraphics[width=\linewidth,height=1.2in]{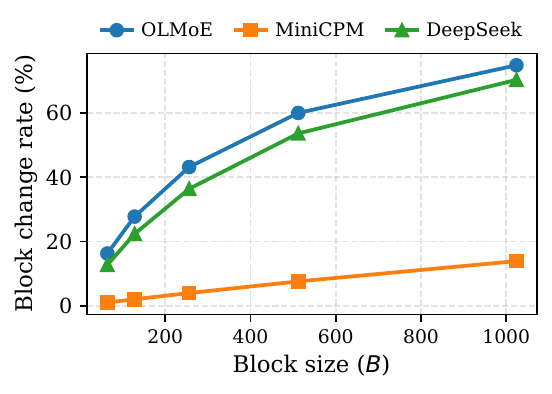}
    \caption{Block change rate.}
    \label{fig:block-level}
  \end{subfigure}
  \caption{Shard-level and block-level exponent change rate.}
  \label{fig:shard-vs-block}
  \vspace{-0.15in}
\end{wrapfigure}
A ZeRO-3 shard can be reused only if each of its $n{\approx}10^6$ exponent values stays unchanged. With per-element change rate $c{=}0.003$ (Section~\ref{sec:exponent-stability}), a uniform per-element model gives $\Pr(\text{shard unchanged}) = (1-c)^n \approx e^{-3000} \approx 0$. Fig.~\ref{fig:shard-level} shows the measured shard change rate versus the world size $W$. The shard/block change rate is the fraction of shards/blocks that contain at least one weight whose exponent changes between consecutive iterations. The world size $W$ is the number of GPUs used for sharding, so each GPU stores $1/W$ of each parameter. The $W{=}4$ point is measured directly and the $W\!\in\!\{8,16,32\}$ points are projected by partitioning the measured per-shard exponent changes across the corresponding number of GPUs. After each parameter update, the shard change rate is high and nearly flat across world sizes, ranging from $59\%$ for MiniCPM to $98.5\%$ for DeepSeek. The fully unchanged fraction runs from about $1.5\%$ for DeepSeek to $41\%$ for MiniCPM, far above the near-zero prediction of the uniform model because exponent changes concentrate in a subset of shard types rather than spreading uniformly. The most stable shards, such as layer norms, stay entirely unchanged, and MiniCPM carries a larger share of these stable shards, which is why its shard change rate is markedly lower. Even so, most shards change every step for every model, so shard-level reuse leaves substantial redundancy unexploited and motivates the finer block-level granularity below.  

To facilitate exponent reuse, we divide each shard into fixed-size blocks of
$B$ weight values each. A block stays reusable with probability $(1{-}c)^B$. 
Fig.~\ref{fig:block-level} shows the block change rate for $B{\in}\{64,128,256,512,1024\}$ across all three models; equivalently, the reuse rate ($1$ minus the change rate) ranges from $84$--$99\%$ at $B{=}64$ down to $25$--$86\%$ at $B{=}1024$. For MiniCPM the change rate grows slowly with $B$, while for OLMoE and DeepSeek it grows sharply.


\vspace{-0.1in}

\begin{observe}\label{obs:block}
Shard-level exponent reuse is limited, as $59$--$99\%$ of shards change each iteration. In contrast, block-level reuse is more effective, with $84$--$99\%$ ($B{=}64$) and $25$--$86\%$ ($B{=}1024$) of blocks remaining reusable across models (Figure~\ref{fig:shard-vs-block}).
\end{observe}
\vspace{-0.0in}

\subsection{Block Change Rate Heterogeneity Across Shards}
\begin{figure*}[t]
    \centering
    \begin{subfigure}[t]{0.32\textwidth} \includegraphics[width=\linewidth]{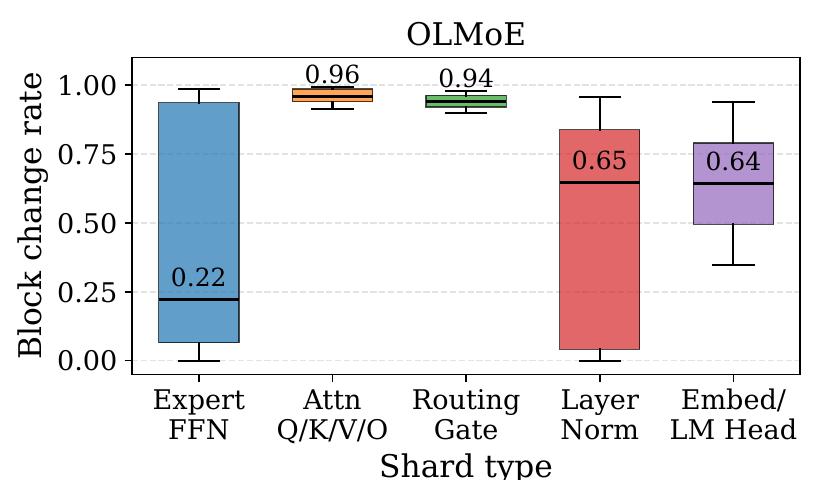}
      \caption{OLMoE-1B-7B}
      \label{fig:block-change-olmoe}
    \end{subfigure}
    \hfill
    \begin{subfigure}[t]{0.32\textwidth}         \includegraphics[width=\linewidth]{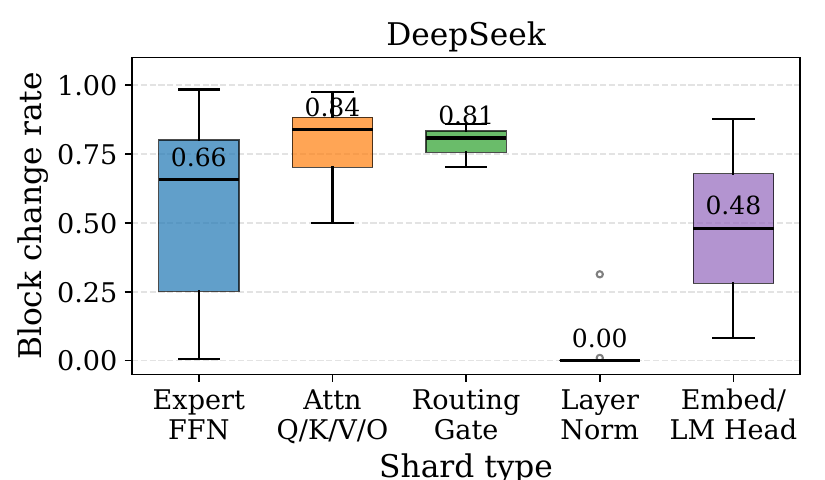}
      \caption{DeepSeek-MoE-16B} \label{fig:block-change-deepseek}
    \end{subfigure}   \hfill
    \begin{subfigure}[t]{0.32\textwidth}
      \includegraphics[width=\linewidth]{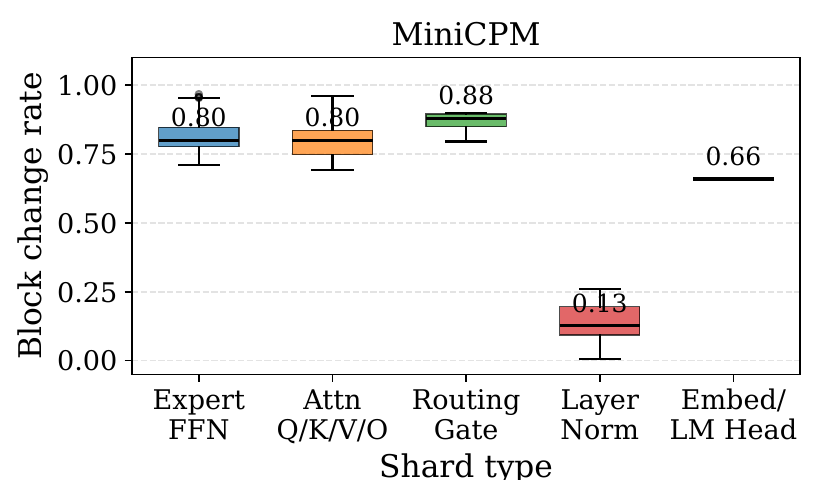} \caption{MiniCPM-MoE-8$\times$2B}
      \label{fig:block-change-minicpm}  \end{subfigure} 
      \caption{
      Block change rate.}
    \label{fig:block-change-by-type} \vspace{-0.25in}
  \end{figure*}

We group shards into five categories: expert FFN, attention, routing gate,
layer norm, and embedding/LM head.
Figure~\ref{fig:block-change-by-type} shows the distribution of block change rate for each
category at block size $B{=}512$.
In OLMoE-1B-7B, expert FFN shards have the lowest median block
change rate at $0.22$. 
Attention and routing shards show much higher medians
of $0.96$ and $0.94$, while layer norm and
embedding/LM head shards fall in between at $0.64$-$0.65$.
In DeepSeek-MoE-16B and MiniCPM-MoE-8$\times$2B, layer norm is the most stable category (medians $0.00$ and $0.13$), while attention, routing-gate, and expert FFN medians sit in the $0.66$--$0.84$ range, leaving only the layer-norm shards reliably reusable.
\vspace{-0.1in}
\begin{observe}\label{obs:blockwise}
Block change rates vary widely across shard types within a model: layer norm is consistently the most stable, while expert FFN is highly stable for OLMoE but less stable for some other models, and attention and routing-gate shards have the highest change rates, so benefits from exponent reuse vary by shard type (Figure~\ref{fig:block-change-by-type}).
\end{observe}


\subsection{Time Overhead}
\begin{wrapfigure}{r}{0.58\columnwidth}
  \vspace{-0.60in}
  \centering
  \begin{subfigure}[t]{0.48\linewidth}
    \includegraphics[width=\linewidth,height=1.2in]{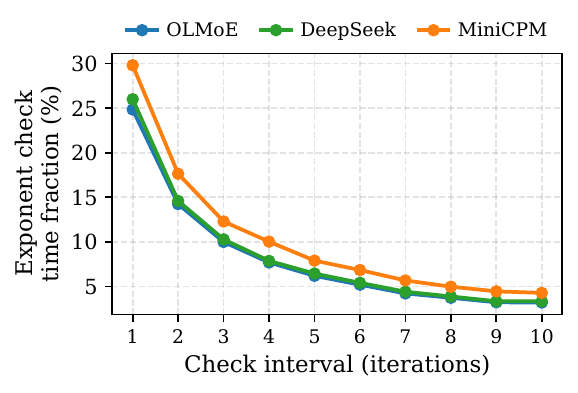}
    \caption{\centering Exponent check time fraction}
    \label{fig:overhead-vs-check}
  \end{subfigure}
  \hfill
  \begin{subfigure}[t]{0.48\linewidth}
    \includegraphics[width=\linewidth,height=1.2in]{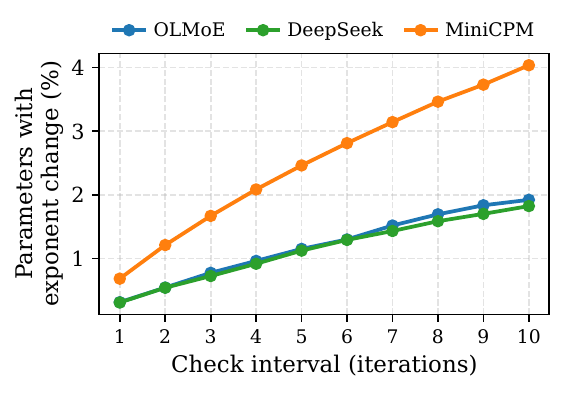}
    \caption{Exponent change rate}
    \label{fig:param-changed-vs-check}
  \end{subfigure}\vspace{-0.05in}
  \caption{Exponent check time and fraction of exponents that changed between checks.}
  \label{fig:exponent-check-overhead}
  \vspace{-0.1in}
\end{wrapfigure}

We measure the runtime overhead of exponent checking by running it on the same
CUDA stream as training computation, without overlapping it with forward,
backward, or communication operations. We measure the exponent check time fraction in the iteration time including the check time. Check interval denotes the number of iterations between successive exponent change checks. Figure~\ref{fig:exponent-check-overhead}(a) shows the exponent check time fraction for all three models. Checking every iteration adds 25--30\% overhead on average, while checking every 10 iterations reduces the overhead to 3--4\%. Figure~\ref{fig:exponent-check-overhead}(b) shows that with interval~1, only 0.3--0.7\% of exponents change on average.
\vspace{-0.1in}
\begin{observe}\label{obs:overhead}
When the exponent check is performed every iteration without operation overlap, exponent checking incurs a 25--30\% time overhead
(Figure~\ref{fig:exponent-check-overhead}).
\end{observe}


\section{System Design}
\begin{wrapfigure}{r}{0.45\columnwidth}
  \vspace{-0.60in}
  \centering
  \includegraphics[width=0.45\columnwidth,height=1.3in]{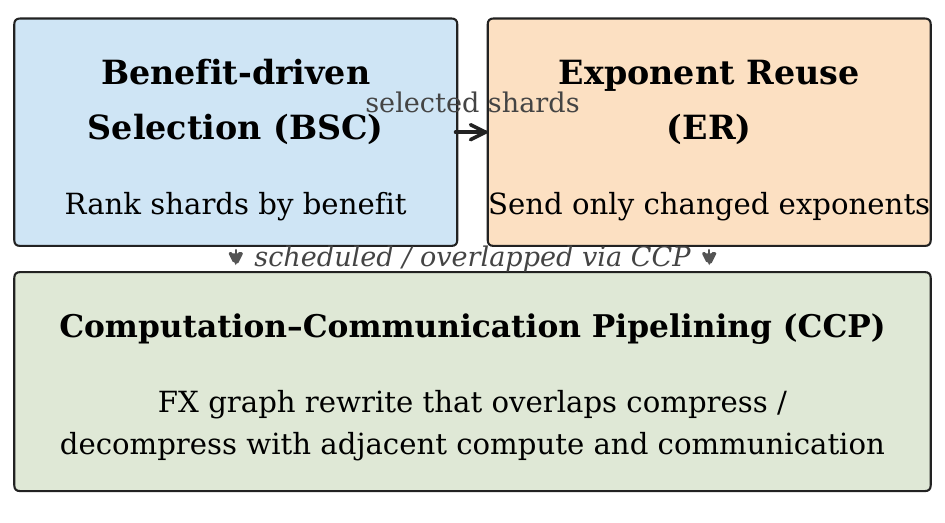}  \vspace{-0.25in}
  \caption{Architecture overview of CIERA.  }
  \label{fig:architecture_overview}
  \vspace{-0.0in}
\end{wrapfigure}
Based on our observations from Section~\ref{sec:motivation}, we design three key components, as illustrated in Figure~\ref{fig:architecture_overview}.
As shown in Figure~\ref{fig:architecture_overview}, after model initialization and warmup, CIERA applies \textbf{exponent reuse based compression} to each selected shard when it reaches the Allgather path, splitting each value into exponent and sign-plus-mantissa and reusing cached exponents when unchanged. 
The set of shards that go through this path is chosen once by \textbf{benefit-driven selective compression}, which estimates the expected communication savings of different parameter shards, ranks them by benefit-to-memory ratio, and selects the highest-ranked shards for exponent caching. This shard selection is performed once and then reused across the following training iterations. 
Finally, CIERA uses \textbf{computation–communication pipelining} to overlap the compression operations with GPU compute operations to hide compression time overhead.

\subsection{Exponent Reuse based Compression}
\label{sec:exponent_reuse}

In floating-point formats such as FP16 and BF16, each value is encoded as sign, exponent, and mantissa bits. The exponent changes only when the magnitude crosses a power-of-two boundary, so it changes much less often than the mantissa during training. This matches our Observation~\ref{obs:stability}, where the exponent change rate stays below 1\% for most iterations. Therefore, repeatedly transmitting unchanged exponent bits is redundant and can instead be reused locally on GPUs. To further reduce communication overhead, we compress the exponents that are changed using ANS via nvCOMP~\cite{nvidia_nvcomp_cpp_api}.



As shown in Observation~\ref{obs:block}, a shard-level reuse decision rarely triggers in practice. To make reuse practical, we check and reuse exponents at a finer {block level} inside each shard. Algorithm~\ref{alg:adera-core} (Appendix) 
illustrates the sender-side compress operations (split, hash, ANS-encode) and receiver-side decompress operations (ANS-decode, merge) for  
one Allgather of shard~$i$. We partition each shard into $S_b^i$ fixed-size blocks. Each GPU, as a sender, maintains a persistent per-block structure: $\mathsf{HashCache}$, and as a receiver, maintains $\mathsf{ExpCache}$. The sender splits each weight value into exponent bits and sign-plus-mantissa bits, and computes its 128-bit hash. In order to reduce collision probability, the 128-bit hash is built from two independent 64-bit SplitMix64 hashes~\cite{steele2014fast,thorup2015high}. 
The hash value of block $k$ in shard $i$ is stored in $\mathsf{HashCache}[i,k]$. We treat a block's exponents as unchanged only when its hash matches the previous hash value. Assuming the two 64-bit SplitMix64 hashes are independent, the per-comparison collision probability is approximately $2^{-128}$; we observed no collisions across all our runs.

The sender creates a bitmap for each shard indicating whether each block has changed; if yes, the sender sets the block's bit in the bitmap and updates its hash in $\mathsf{HashCache}$.  The sender sends compressed exponents that are changed, and the sign-plus-mantissa values of all shard blocks, along with the bitmap to the receiver. The communication amount is reduced by not sending the unchanged exponents and compressing the changed exponents.



After the receiver receives the data, it obtains the positions of changed blocks from the bitmap. 
It reads the previous exponents of all shard blocks from $\mathsf{ExpCache}[i,\cdot]$, ANS-decodes the exponents of changed blocks, replaces only the changed blocks with the decompressed blocks, and reconstructs the full shard by merging all the exponents with the received sign-plus-mantissa values. 
The hash, ANS encode, and ANS decode steps run on a separate CUDA stream so they overlap with communication and computation using our Computation–Communication Pipelining in ~\ref{sec:graph_scheduling}.

\subsection{Benefit-driven Selective Compression}
\label{sec:partial_compression}

Based on Observation~\ref{obs:blockwise}, \name compresses only shards where compression actually pays off. The gain formula (next paragraph) ranks shards by the ratio of expected byte savings to extra cache memory; in practice this selects shards with low block change rates (predominantly layer norm and expert FFN) and skips shards with high change rates (attention, routing gate) because compression cannot pay for itself when almost every block changes.


For the targeted shards, \name ranks the shards using a gain formula that takes each shard's offline-measured block change rate $q_i$ as input. A shard with a low $q_i$ has a large gain and is compressed. A shard with a high $q_i$ has a zero or negative gain and is not compressed. Shard~$i$ has $N^i$ weight values in $b$-bit precision (16 for BF16/FP16), so it occupies $S^i = N^i\,b/8$ bytes, of which $S_e^i$ are exponent bytes and $S_{s+m}^i = S^i - S_e^i$ are sign-plus-mantissa. We partition each shard into blocks of $B$ values, giving $S_b^i = \lceil N^i / B \rceil$ blocks and a one-bit-per-block bitmap of $S^i_{\text{bm}} = \lceil S_b^i / 8 \rceil$ bytes. 

    Among the target shards, \name computes a per-iteration latency gain $g_i$ at initialization and keeps only shards with positive gain:
$g_i =
\mathrm{Comm}(S^i)
-
\left(
\mathrm{Comm}(S_{s+m}^i + S_{\text{bm}}^i + q_i S_e^i)
+ T^i_{\text{hash}}
+ T^i_{\text{cmp}}
+ T^i_{\text{decmp}}
\right)$, where $\mathrm{Comm}(\cdot)$ maps a payload size to the measured Allgather latency (profiled at initialization), $T^i_{\text{hash}}$ is the total block hashing time of the shard, $T^i_{\text{cmp}}$ is the sender-side time for compressing, and $T^i_{\text{decmp}}$ is the receiver-side time for decompressing the received exponents and reconstructing the full shard by merging them with the cached exponents and sign-plus-mantissa bytes.

Using Computation--Communication Pipelining in Section~\ref{sec:graph_scheduling}, $T^i_{\text{hash}}$ and $T^i_{\text{cmp}}$ are effectively zero on the critical path because these kernels execute on a dedicated CUDA stream that runs concurrently with the Allgather.
Only $T^i_{\text{decmp}}$ remains on the critical path because reconstruction runs after the Allgather completes.

With cache footprint $M_i = (W-1)\,S_e^i$ per GPU at world size $W$, we rank shards in descending order of $g_i/M_i$ and select them until the cache budget is exhausted. The selected set is fixed for the training run. At runtime, only blocks whose hash differs from the previous iteration are transmitted, so the expected payload per iteration is $S_{s+m}^i + S_{\text{bm}}^i + q_i S_e^i$ bytes.
\vspace{-0.1in}

\subsection{Computation--Communication Pipelining}
\label{sec:graph_scheduling}

Compression and decompression operations add latency that can exceed the communication savings if executed sequentially. We address this by scheduling these operations early in the execution workflow to overlap them with preceding compute kernels and data transfers. 

We use a PyTorch FX graph~\cite{reed2022torch} to represent the model’s execution. In the original graph, each shard $S$ in layer $L_i$ is associated with an Allgather node $\mathrm{Allgather}(L_i)$, which gathers the layer’s parameter shards from all GPUs. We insert four additional kernel nodes for each parameter shard chosen by benefit-driven selective compression: 1) the $\mathrm{Hash}(L_i)$ node computes a hash over exponent bits to detect changes; 2) the $\mathrm{PreDecmp}(L_i)$ node prepares the receiver's cached exponents for reconstruction; 3) the $\mathrm{Decmp}(L_i)$ node decompresses the received exponents in that iteration for reconstruction; and 4) the $\mathrm{Cmp}(L_i)$ node builds the Allgather payload, packaging either compressed exponents plus sign-plus-mantissa bits when exponents changed, or only sign-plus-mantissa bits when exponents are reused.

\begin{figure}[t]
  \centering
  \includegraphics[width=\linewidth]{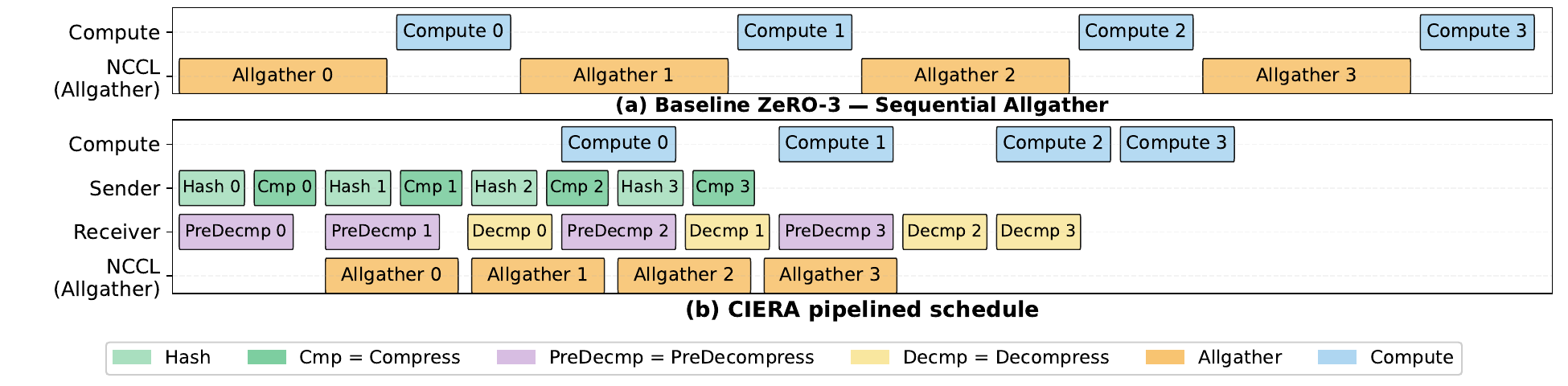}
  \caption{\centering Graph scheduling for overlapping communication with computation.}
  \label{fig:graph_scheduling}
  \vspace{-0.26in}
\end{figure}

\DEL{These extra inserted kernels (Hash, PreDecompress, Decompress, and Compress) can increase iteration time if they run in sequence right before sending the shard or right before the shard is used. We avoid this by moving these operations earlier in the schedule as possible, so they run concurrently with the compute and communication of earlier layers.}


Figure~\ref{fig:graph_scheduling} illustrates the transformation of the workflow. Figure~\ref{fig:graph_scheduling}(a) shows the baseline ZeRO-3 execution, where per-layer compute is followed by its corresponding Allgather in sequence. Figure~\ref{fig:graph_scheduling}(b) shows \name's schedule, where hash, predecompress, and compression operations are inserted for compressed layers and moved earlier so they execute in parallel with the compute of the preceding layer. This overlap effectively masks compression latency with concurrent computation. The subscript $i$ in Figure~\ref{fig:graph_scheduling}(b) indexes one Allgather event in the fixed order $A = [l_1, \ldots, l_m]$, so $\mathrm{Hash}_i$, $\mathrm{PreDecmp}_i$, $\mathrm{Cmp}_i$, $\mathrm{Allgather}_i$, $\mathrm{Decmp}_i$, and $\mathrm{Compute}_i$ are the operations that handle layer $L_i$'s parameter shard in a single training iteration. The four squares $\mathrm{Hash}_0, \mathrm{Hash}_1, \mathrm{Hash}_2, \mathrm{Hash}_3$ on the sender row are the hash kernels for four example layers of the same iteration, not repeated hashes of one layer across four iterations. Each square is one GPU kernel and the rows group kernels by the GPU role or stream that executes them: NCCL Allgather, Receiver-side work, Sender-side work, and Compute.

For each compressed layer $L_i$ (Figure~\ref{fig:graph_scheduling}(b)), the sender runs $\mathrm{Hash}_i$ and $\mathrm{Cmp}_i$ to detect changed blocks and pack the payload, while the receiver runs $\mathrm{PreDecmp}_i$ over its local cache in parallel because it has no cross-GPU dependency. $\mathrm{Allgather}_i$ transmits the payload, then $\mathrm{Decmp}_i$ replaces the changed blocks in the cached exponents and combines them with the received sign-plus-mantissa to reconstruct the layer for $\mathrm{Compute}_i$. Since $\mathrm{Hash}_{i+1}$, $\mathrm{Cmp}_{i+1}$, and $\mathrm{PreDecmp}_{i+1}$ can begin while $\mathrm{Allgather}_i$ is still in flight, only $\mathrm{Decmp}_i$ remains on the critical path before $\mathrm{Compute}_i$ launches.

\section{Performance Evaluation}
\label{sec:eval}

\subsection{Experimental Settings}
\label{sec:exp-settings}
The setup follows Section~\ref{sec:motivation}, extended with two additional MoE models:  Mixtral-8$\times$7B~\cite{jiang2024mixtral} (8 experts), and Llama-4-Scout-17B-16E~\cite{meta2025llama4} (16 experts). Adam~\cite{kingma2014adam} optimizer is used by default. Llama-4-Scout-17B-16E does not fit at $L{=}4$ on 4 GPUs with Adam, so we use Adafactor~\cite{shazeer2018adafactor} for that model. We enable activation checkpointing~\cite{zhao2023pytorch,maurya2024deep} at Transformer-layer granularity in DeepSpeed 0.17.2. Multi-node experiments run on nodes of 8 A100-80GB GPUs, with 200~Gbps HDR InfiniBand between nodes (measured cross-node AllGather bus bandwidth ${\approx}8$~GB/s) and 600~GB/s NVLink (NVSwitch) within a node. The 8-GPU configuration is single-node and the 16-GPU configuration spans two nodes. Starred ($\star$) entries in Tables~\ref{tab:lossless-scaling} and~\ref{tab:fp16-results} are measured on this testbed, and unstarred entries are simulator projections. Under FP16 only OLMoE is measured against ZeRO++ on the testbed, so the other FP16 entries such as MiniCPM are projections, whereas under BF16 both OLMoE and MiniCPM are measured against ZeRO-3. We set block size $B{=}512$ empirically, balancing exponent-reuse rate against per-block kernel-launch overhead. We project unmeasured 32--128 GPU cases with a trace-driven simulator. We collect per-layer iteration traces for all six models at $L\in\{2,4,6,8\}$ on 4 and 8 A100 GPUs and anchor the projection on the measured full-model speedups at 4, 8, and 16 GPUs. The simulator matches these measured iteration times within $4\%$ on full OLMoE and MiniCPM at 8 and 16 GPUs.

\subsection{Comparison Against Lossless Baselines}
\vspace{-0.1in}
\label{sec:eval-lossless}
\begin{wrapfigure}{r}{0.62\columnwidth}
  \vspace{-0.25in}
  \centering
  \begin{subfigure}[t]{0.48\linewidth}
    \includegraphics[width=\linewidth,height=1.2in]{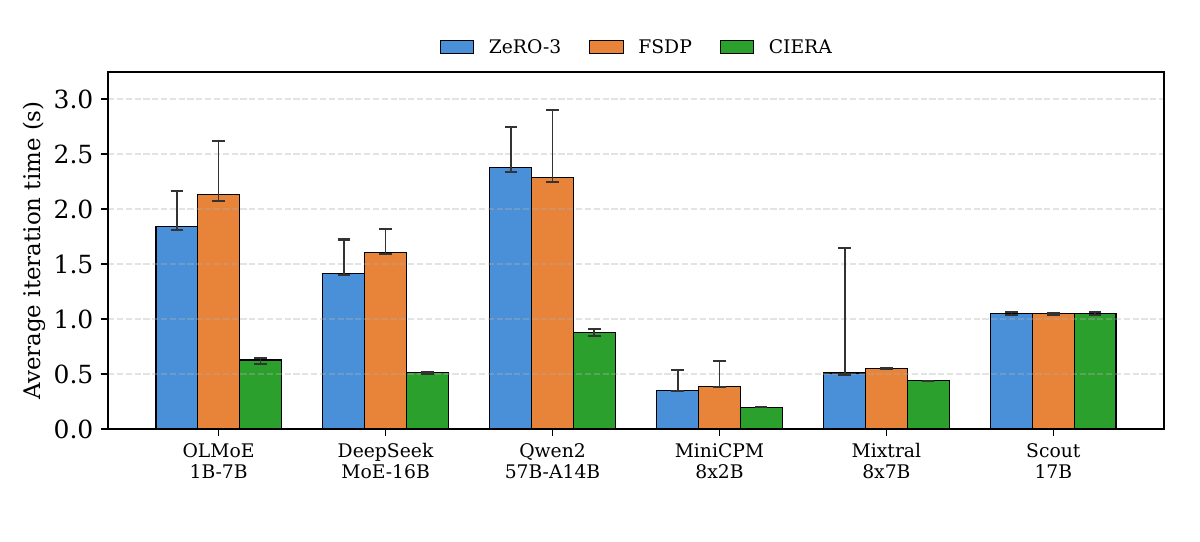}
    \caption{\centering CIERA against lossless baselines.}
    \label{fig:result-lossless}
  \end{subfigure}
  \hfill
  \begin{subfigure}[t]{0.48\linewidth}
    \includegraphics[width=\linewidth,height=1.2in]{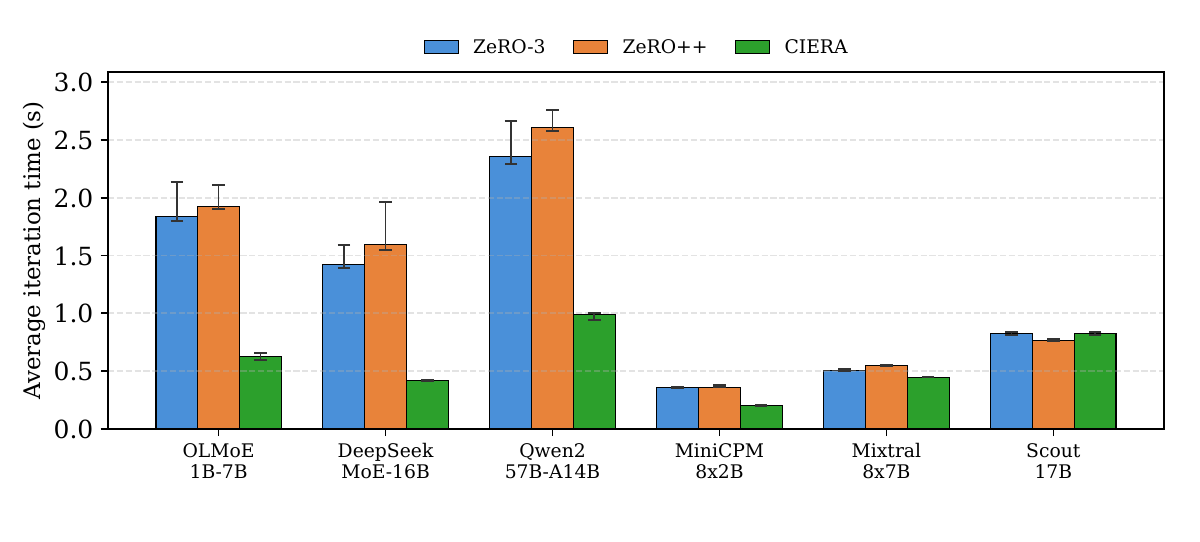}
    \caption{\centering CIERA against lossy baseline}
    \label{fig:result-fp16-4gpu}
  \end{subfigure}
   \caption{\centering Performance of CIERA.}
  \vspace{-0.15in}
\end{wrapfigure}
We compare \name against two lossless baselines: DeepSpeed ZeRO-3~\cite{rajbhandari2020zero} and PyTorch FSDP~\cite{zhao2023pytorch}, both of which transmit full-precision parameters without compression. Figure~\ref{fig:result-lossless} shows the per-model iteration time on 4 GPUs with error whiskers marking the 5th and 95th percentile over 1000 post-warmup iterations. \name achieves $1.16$--$2.89\times$ iteration-time speedup over ZeRO-3 on the five communication-bound models and up to $3.38\times$ speedup over FSDP on OLMoE (higher speedup is better). Llama-4-Scout-17B stays at $1.00\times$ on 4 GPUs because BSC selects no shards: high per-shard change rates plus fast intra-node NVLink keep every shard's profiler score $g_i$ non-positive; Scout becomes profitable once inter-node bandwidth dominates, reaching $1.16\times$ at 32 GPUs (Table~\ref{tab:lossless-scaling}). Table~\ref{tab:lossless-scaling} extends to 8--128 GPUs (real measurements marked $\star$, others projected by the simulator), reaching $1.16$--$4.28\times$ iteration-time speedup over ZeRO-3 as inter-node communication takes a larger share of iteration time.

\begin{table*}[t]
\centering
\begin{minipage}[t]{0.49\textwidth}
\centering
\caption{Speedup of \name over ZeRO-3 (BF16, $S{=}1024$,
         full model layers with Adam; Adafactor for Llama-4-Scout).}
\label{tab:lossless-scaling}
\small
\resizebox{\linewidth}{!}{%
\begin{tabular}{lccccc}
\toprule
\textbf{Model/GPU\#}
  & \textbf{8} & \textbf{16}
  & \textbf{32} & \textbf{64} & \textbf{128} \\
\midrule
OLMoE-1B-7B         & 3.70$\star$ & 3.70$\star$ & 3.89 & 4.08 & 4.28 \\
MiniCPM-8$\times$2B & 1.38$\star$ & 2.12$\star$ & 2.23 & 2.34 & 2.45 \\
DeepSeek-MoE-16B    & ---         & 3.34        & 3.51 & 3.68 & 3.87 \\
Qwen2-57B-A14B      & ---         & 2.66        & 2.79 & 2.93 & 3.08 \\
Mixtral-8$\times$7B & ---         & 1.33        & 1.40 & 1.47 & 1.54 \\
Llama-4-Scout-17B   & ---         & ---         & 1.16 & 1.22 & 1.28 \\
\bottomrule
\end{tabular}
}
\end{minipage}
\hfill
\begin{minipage}[t]{0.49\textwidth}
\centering
\caption{Speedup of \name over ZeRO++ (FP16, $S{=}1024$,
         full model layers with Adam).}
\label{tab:fp16-results}
\small
\resizebox{\linewidth}{!}{%
\begin{tabular}{lccccc}
\toprule
\textbf{Model/GPU\#}
  & \textbf{8} & \textbf{16}
  & \textbf{32} & \textbf{64} & \textbf{128} \\
\midrule
OLMoE-1B-7B         & 3.93$\star$ & 3.68$\star$ & 4.02 & 4.22 & 4.42 \\
MiniCPM-8$\times$2B & 1.74        & 2.60        & 2.73 & 2.87 & 3.01 \\
DeepSeek-MoE-16B    & ---         & 1.93        & 2.03 & 2.13 & 2.24 \\
Qwen2-57B-A14B      & ---         & 3.08        & 3.23 & 3.40 & 3.57 \\
Mixtral-8$\times$7B & ---         & 1.67        & 1.75 & 1.84 & 1.93 \\
Llama-4-Scout-17B   & ---         & ---         & 2.03 & 2.13 & 2.24 \\
\bottomrule
\end{tabular}
}
\end{minipage}
\vspace{-0.15in}
\end{table*}

\vspace{-0.1in}

\subsection{Comparison Against Lossy Baseline}
\label{sec:eval-lossy}
\vspace{-0.1in}

We compare \name against ZeRO++~\cite{wang2023zero++}, a lossy baseline that uses int8 Allgather for parameters, int4 ReduceScatter for gradients, and per-node parameter replicas. ZeRO++ requires FP16, so all systems run under FP16 here. Other lossy systems are not directly comparable on sharded MoE: SDP4Bit~\cite{jia2024sdp4bit} is dense-LLM only, gZCCL~\cite{huang2024gzccl} is an MPI/MVAPICH framework rather than a NCCL/DeepSpeed plug-in, and TAGC~\cite{polyakov2025tagc} compresses gradients on the ReduceScatter path; we discuss them in Section~\ref{sec:related}.

Figure~\ref{fig:result-fp16-4gpu} compares the average iteration time on 4 GPUs with FP16 ($L{=}4$ layers, $S{=}1024$, six MoE models grouped on the x-axis with three bars per model for ZeRO-3 / ZeRO++ / \name). \name achieves $1.13$--$3.38\times$ iteration-time speedup over ZeRO-3 and $1.21$--$3.80\times$ iteration-time speedup over ZeRO++ on the five communication-bound models (higher speedup is better), while Scout stays at $1.00\times$, same iteration time as the baseline, due to zero-shard selection. Table~\ref{tab:fp16-results} extends the comparison to 8--128 GPUs and reaches $1.67$--$4.42\times$ iteration-time speedup over ZeRO++. On OLMoE the speedup over ZeRO++ is higher at 8 GPUs ($3.93\times$) than at 16 ($3.68\times$), which reflects the ZeRO++ baseline rather than \name. Within a single node NVLink is not the bottleneck, so ZeRO++'s weight quantization is almost pure overhead and its iteration time stays high. Across nodes the same quantization reduces inter-node traffic and speeds ZeRO++ up, which narrows \name's relative margin. Against the unquantized ZeRO-3 baseline (Table~\ref{tab:lossless-scaling}), OLMoE is communication-bound at both 8 and 16 GPUs, so its speedup is flat at $3.70\times$. The scaling model fits a monotone log-linear trend through the measured 4/8/16-GPU anchors, so the 32--128-GPU projections rise past these close or slightly non-monotone measured points.




\subsection{Ablation Study}
\label{sec:eval-ablation}

We evaluate the incremental contribution of \name components by progressively
enabling them on top of ZeRO-3 across all six MoE models
($L{=}4$, $S{=}1024$, 4~A100 GPUs, BF16). ER is executed through our
GraphATen rewrite of ZeRO-3's gather/release scheduling, which provides the
execution path required for compressed AllGather. The subsequent columns then
add BSC and CCP incrementally, so every column's speedup is cumulative over ZeRO-3.
Table~\ref{tab:ablation} reports the per-iteration time of ZeRO-3 and the
cumulative speedup of \name components.

\begin{wraptable}{r}{0.50\columnwidth}
\vspace{-0.10in}
\centering
\caption{Ablation study on 4 A100 GPUs ($L{=}4$, $S{=}1024$, BF16).}
\label{tab:ablation}
\small
\resizebox{\linewidth}{!}{%
\begin{tabular}{lcccc}
\toprule
\textbf{Model}
  & \textbf{ZeRO-3}
  & \textbf{+ER}
  & \textbf{+BSC}
  & \textbf{+CCP} \\
\midrule
OLMoE-1B-7B         & 1.840s & 1.76$\times$ & 2.84$\times$ & 2.89$\times$ \\
DeepSeek-MoE-16B    & 1.412s & 1.70$\times$ & 2.82$\times$ & 2.86$\times$ \\
Qwen2-57B-A14B      & 2.377s & 1.69$\times$ & 2.66$\times$ & 2.75$\times$ \\
MiniCPM-8$\times$2B & 0.349s & 1.27$\times$ & 1.74$\times$ & 1.75$\times$ \\
Mixtral-8$\times$7B & 0.508s & 0.96$\times$ & 1.15$\times$ & 1.16$\times$ \\
Llama-4-Scout-17B   & 1.053s & 1.00$\times$ & 1.00$\times$ & 1.00$\times$ \\
\bottomrule
\end{tabular}
}
\vspace{-0.15in}
\end{wraptable}

\textbf{Exponent Reuse (ER) based Compression.}
ER is evaluated using our GraphATen execution path, which enables compressed
AllGather within ZeRO-3's gather/release schedule. The reported +ER configuration
therefore measures exponent reuse together with the execution support required to
realize it, rather than ER as an isolated codec.
When enabled across all shards with no selection, this configuration reaches
$1.27$--$1.76\times$ over ZeRO-3 on OLMoE, DeepSeek, Qwen2, and MiniCPM while sending
only the changed-exponent blocks across iterations. Mixtral, however, runs
$0.96\times$ slower than ZeRO-3: its block change rate is high, so the per-shard
hash and packing overhead exceeds the bytes saved.

\textbf{Benefit-driven Selective Compression (BSC).}
Adding the benefit-driven profiler on top of ER skips shards where compression does not pay off. The largest gain is on Mixtral: BSC removes the unprofitable shards and the speedup goes from $0.96\times$ to $1.15\times$. On the bytes-bound models the speedup also grows: OLMoE $1.76\times{\to}2.84\times$, DeepSeek $1.70\times{\to}2.82\times$, Qwen2 $1.69\times{\to}2.66\times$, MiniCPM $1.27\times{\to}1.74\times$. Llama-4-Scout selects zero shards on 4-GPU NVLink, so it stays at $1.00\times$.

\textbf{Computation-Communication Pipelining (CCP).}
Overlapping hash, compress, and decompress kernels with preceding backward compute adds a small further gain on top of BSC$+$ER ($+0.04$--$0.09\times$ on OLMoE, DeepSeek, and Qwen2; under $1\%$ on MiniCPM and Mixtral, where BSC$+$ER is near its ceiling). Scout stays at $1.00\times$ because BSC selected zero shards, so CCP has nothing to overlap. Each column adds one component on top of the previous configuration, so all numbers are cumulative iteration-time speedups over ZeRO-3. We do not report standalone-component runs (e.g., CCP alone is undefined because there are no compress / decompress kernels to overlap).
\vspace{-0.1in}
\subsection{Sensitivity Analysis}
\vspace{-0.1in}
\label{sec:eval-sensitivity}

\begin{figure}[t]
  \centering
  \begin{subfigure}[t]{0.32\columnwidth}
    \includegraphics[width=\linewidth,height=1.2in]{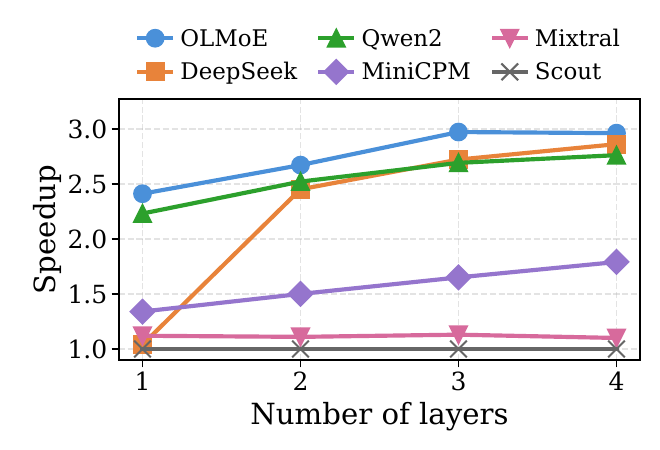}
    \caption{Speedup vs.\ \# of layers}
    \label{fig:sens-layers}
  \end{subfigure}
  \hfill
  \begin{subfigure}[t]{0.32\columnwidth}
    \includegraphics[width=\linewidth,height=1.2in]{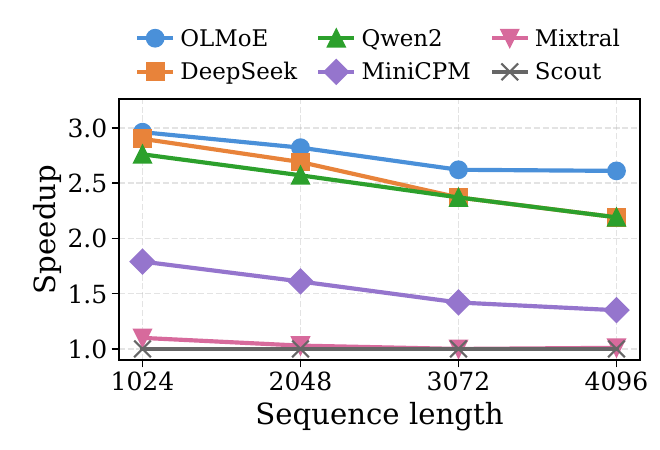}
    \caption{Speedup vs.\ sequence length}
    \label{fig:sens-seqlen}
  \end{subfigure}
  \hfill
  \begin{subfigure}[t]{0.32\columnwidth}
    \includegraphics[width=\linewidth,height=1.2in]{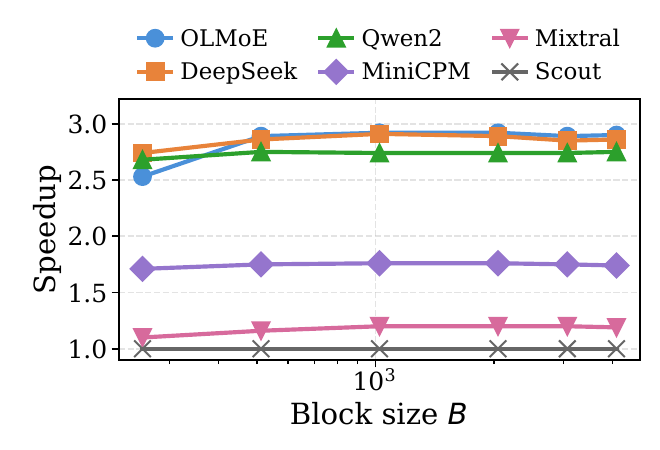}
    \caption{Speedup vs.\ block size }
    \label{fig:sens-blocksize}
  \end{subfigure}
  \caption{Sensitivity of \name speedup over ZeRO-3 to number of transformer
           layers,  sequence length, and block size.}
  \label{fig:sensitivity}
  \vspace{-0.25in}
\end{figure}

%


\noindent\textbf{Layer count.}
Figure~\ref{fig:sens-layers} varies the number of layers, $L$, from 1 to 4 at sequence length, $S{=}1024$. Iteration-time speedup over ZeRO-3 (higher is faster) generally rises with more layers because of more Allgather calls: OLMoE, DeepSeek, Qwen2, and MiniCPM all improve significantly, with DeepSeek showing the largest gain (from $1.04\times$ to $2.86\times$ iteration-time speedup over ZeRO-3). Mixtral stays near $1.16\times$ iteration-time speedup, and Scout stays at $1.00\times$ (same iteration time as ZeRO-3) because the profiler selects zero shards on NVLink.

\noindent\textbf{Sequence length.}
Figure~\ref{fig:sens-seqlen} varies sequence length $S$ from 1024 to 4096 at $L{=}4$. As $S$ grows, compute increases while Allgather volume is fixed, so the relative benefit of compression shrinks: the iteration-time speedup over ZeRO-3 falls but stays above $1\times$, e.g., OLMoE from $2.89\times$ to $2.61\times$, MiniCPM from $1.75\times$ to $1.35\times$.


\noindent\textbf{Block size.}
Figure~\ref{fig:sens-blocksize} varies $B$ from 256 to 4096 at $L{=}4$, $S{=}1024$. Each block has a 128-bit hash kept locally on the sender and contributes one bit to the per-shard change bitmap that travels with every Allgather. Smaller $B$ means more blocks per shard, which inflates the number of per-block kernel launches (hash, ANS-encode, ANS-decode) and slightly enlarges the bitmap. Larger $B$ reduces these per-block kernel launches but each block is more likely to contain at least one changed exponent, which lowers reuse. $B{=}512$ balances the two and gives the headline speedups in Section~\ref{sec:eval-lossless}. At $B{=}256$ each model loses a few percent of iteration time to the extra per-block kernel launches, and from $B{=}1024$ on the curves stay within a few percent of $B{=}512$. Scout stays at $1.00\times$ as discussed earlier.

\subsection{Correctness Evaluation}
\label{sec:correctness}
\vspace{-0.1in}

\begin{wrapfigure}{r}{0.40\columnwidth}
  \vspace{-0.3in}
  \begin{subfigure}[t]{0.40\columnwidth}
    \includegraphics[width=\linewidth,height=1.2in]{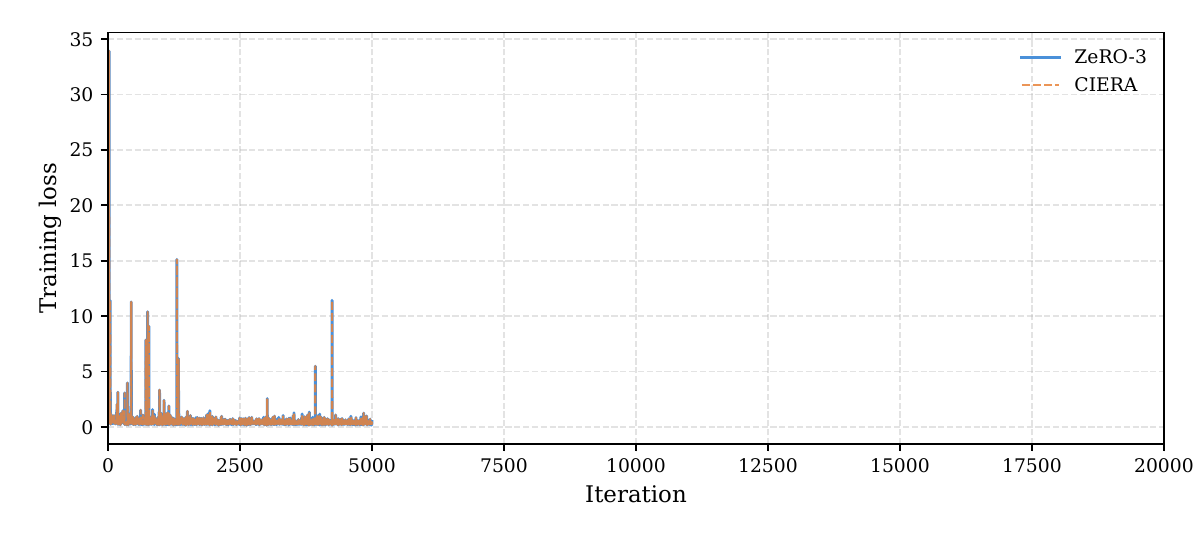}
    
  \end{subfigure}
  \caption{\centering Training loss for OLMoE-1B-7B ($L{=}4$, $S{=}1024$, BF16)}
              \label{fig:loss-convergence}
  \vspace{-0.10in}
\end{wrapfigure}
We verify that \name produces bitwise identical training dynamics to
uncompressed ZeRO-3. Both runs on OLMoE-1B-7B. Figure~\ref{fig:loss-convergence} shows the training loss of
OLMoE-1B-7B over 20{,}000 iterations. The two curves overlap
exactly: across all 20{,}000 steps, the maximum absolute difference
in loss is \textbf{zero}. Every gathered parameter is reconstructed
bit-for-bit at each Allgather boundary, confirming that exponent
caching and block-level reuse introduce no numerical error. 

\section{Related Work}
\label{sec:related}
\vspace{-0.1in}

\noindent\textbf{Gradient compression and sparsification.}
Early work reduces backward traffic through quantization or sparsification: 1-bit SGD and QSGD for low-bit gradients, Top-k or threshold rules that drop small entries~\cite{alistarh2017qsgd,aji2017sparse}, Deep Gradient Compression's momentum correction and error feedback~\cite{lin2017deep}, and adaptive variants motivated by limited gains from fixed levels~\cite{agarwal2022utility}. TAGC targets transformer gradients with format-aware lossless compression on selected layers~\cite{polyakov2025tagc}.

\noindent\textbf{Parameter compression and quantization.}
Recent systems compress parameters during the forward path. ZeRO++~\cite{wang2023zero++} integrates low-bit weight exchange inside FSDP. SDP4Bit~\cite{jia2024sdp4bit} exploits cross-iteration redundancy by lossily quantizing $\Delta W = W_t - W_{t-1}$ to four bits, but treats each weight as a single number with no field-level decomposition. NeuZip~\cite{hao2024neuzip} compresses weights by entropy to reduce on-device memory. ZipServ~\cite{fan2026zipserv} fuses a hardware-aware lossless format into the GEMM kernel to accelerate LLM \emph{inference} weight loading. 

\noindent\textbf{System-aware communication.}
Prior work reduces distributed training communication by adapting compression, reducing collective traffic, or rescheduling communication with computation. Adaptive gradient compression methods such as AdaCGD~\cite{makarenko2022adaptive}, GraVAC~\cite{tyagi2023gravac}, Kimad~\cite{xin2023kimad}, and Accordion~\cite{agarwal2020accordion} tune compression based on signal quality or bandwidth, but focus on gradient communication through ReduceScatter or AllReduce. System-level approaches such as Cupcake~\cite{wang2023cupcake}, OmniReduce~\cite{fei2021efficient}, SqueezeNIC~\cite{rebai2024squeezenic}, and gZCCL~\cite{huang2024gzccl} reduce collective communication cost through traffic reduction or codec and network overlap. Other systems address related communication bottlenecks, including LSH-MoE~\cite{nie2024lshmoe} for MoE all-to-all token dispatch and Comet~\cite{comet2025} or FlashOverlap~\cite{flashoverlap2026} for compute-communication rescheduling. Closest to us, ZipCCL~\cite{lin2026zipcclefficientlosslessdata} losslessly codes BF16 exponents by exploiting the concentrated exponent distribution of approximately Gaussian LLM tensors, without exploiting cross-iteration exponent reuse, while UCCL-Zip~\cite{ma2026ucclzip} integrates lossless compression into NCCL collectives and P2P transfers and evaluates it for RL weight synchronization and distributed LLM inference.

\section{Conclusion}
\vspace{-0.1in}
We observe that exponent values remain stable across most iterations of MoE training, and thus propose CIERA, which reuses cached exponents for compression to reduce Allgather communication overhead, while selectively compressing only those shards where the benefits outweigh the associated costs. The key contribution of \name{} is enabling bit-exact, lossless compression with low decompression overhead. Also, \name{} overlaps compression with computation and communication to hide runtime cost. Our experiments demonstrate that \name{} outperforms the evaluated baselines on communication-bound MoE models. 

\textbf{Limitations and future work.}
\name targets sparse MoE training where exponent stability gives a large pool of compressible shards. On dense LLMs the same selection rule prunes most shards and the speedup collapses. Beyond 16 GPUs we use a trace-driven simulator anchored on real 4-, 8-, and 16-GPU measurements (within 4\% on 8--16 GPUs). The 32-GPU results are short-range extrapolations close to the measured range, while 64- and 128-GPU results require more caution. Multi-node measurements at 32+ GPUs and an online controller for BSC are future work.

\textbf{Broader impacts.}
\name can reduce the communication cost of sharded MoE training, which may lower GPU time, energy use, and training cost for large models. By preserving bitwise-exact parameter reconstruction, it avoids the accuracy risks introduced by lossy communication compression. This can help researchers train larger models under limited hardware budgets. At the same time, making large-scale training more efficient may also lower the marginal cost of scaling, potentially increasing overall demand for computation; thus, the net environmental and economic impact depends on how the resulting efficiency gains are ultimately utilized.
\bibliography{bibliography}
\bibliographystyle{unsrt}

\newpage
\appendix

\section{Block-level exponent reuse algorithm}
\label{app:adera-core-alg}

\begin{algorithm}[H]
\caption{\name\ block-level exponent reuse for one Allgather of shard $i$.}
\label{alg:adera-core}
\KwIn{shard $i$ with $S_b^i$ blocks of $\overline{n_{\text{b}}}$ values; persistent $\mathsf{HashCache}[i,\cdot]$ on sender; persistent $\mathsf{ExpCache}[i,\cdot]$ on receiver.}

\tcp{\textbf{Sender}}
$(\text{exp},\text{mant}) \leftarrow \textsc{SplitExpMant}(\text{shard}_i)$\;
$\text{bitmap}\leftarrow [0]^{S_b^i}$;\quad $\text{payload}\leftarrow \emptyset$\;
\For{$k\leftarrow 1$ \KwTo $S_b^i$}{
  $h\leftarrow\textsc{Hash128}(\text{exp}[k])$\;
  \If{$h\neq \mathsf{HashCache}[i,k]$}{
    $\text{bitmap}[k]\leftarrow 1$;\quad $\mathsf{HashCache}[i,k]\leftarrow h$\;
  }
}
\If{any bit in $\text{bitmap}$ is set}{
  $\text{payload}\leftarrow\textsc{ANSEncode}\!\bigl(\{\text{exp}[k]\mid \text{bitmap}[k]{=}1\}\bigr)$\;
}
$\textsc{Send}(\text{bitmap},\,\text{mant},\,\text{payload})$\;

\tcp{\textbf{Receiver}}
$(\text{bitmap},\text{mant},\text{payload})\leftarrow\textsc{Recv}()$\;
\If{$\text{payload}\neq\emptyset$}{
  $\text{newExp}\leftarrow\textsc{ANSDecode}(\text{payload})$;\quad $j\leftarrow 0$\;
  \For{$k\leftarrow 1$ \KwTo $S_b^i$}{
    \If{$\text{bitmap}[k]=1$}{
      $\mathsf{ExpCache}[i,k]\leftarrow \text{newExp}[j]$;\quad $j\leftarrow j{+}1$\;
    }
  }
}
$\text{shard}_i\leftarrow\textsc{Merge}(\mathsf{ExpCache}[i,\cdot],\,\text{mant})$\;
\end{algorithm}

\end{document}